\documentclass[prd,nofootinbib,preprint,superscriptaddress]{revtex4}

\usepackage{amsmath, amssymb, amsthm, graphicx, epsfig, fancyhdr,epsfig}

\usepackage{natbib}
\usepackage{float}

\usepackage{tikz,xcolor,hyperref}

\definecolor{lime}{HTML}{A6CE39}
\DeclareRobustCommand{\orcidicon}{
	\begin{tikzpicture}
	\draw[lime, fill=lime] (0,0) 
	circle [radius=0.2] 
	node[white] {{\fontfamily{qag}\selectfont \tiny ID}};
	\draw[white, fill=white] (-0.0625,0.095) 
	circle [radius=0.007];
	\end{tikzpicture}
	\hspace{-2mm}
}

\foreach \x in {A, ..., Z}{\expandafter\xdef\csname orcid\x\endcsname{\noexpand\href{https://orcid.org/\csname orcidauthor\x\endcsname}
			{\noexpand\orcidicon}}
}

\newcommand{\be}{\begin{equation}}
\newcommand{\ee}{\end{equation}}
\newcommand{\bea}{\begin{eqnarray}}
\newcommand{\eea}{\end{eqnarray}}

\begin{document}

\title{Mass Gap in Strongly Coupled Infinite Derivative \\ Non-local Higgs: Dyson-Schwinger Approach }

\author{Marco Frasca\orcidA{}}
\email{marcofrasca@mclink.it}
\affiliation{Rome, Italy}

\author{Anish Ghoshal\orcidB{}}
\email{anish.ghoshal@roma2.infn.it}
\affiliation{INFN - Sezione Roma “Tor Vergata”, Via della Ricerca Scientifica 1, 00133, Roma, Italy}

\begin{abstract}
\textit{
We investigate the non-perturbative degrees of freedom in the class of non-local Higgs theories that have been proposed as an ultraviolet completion 4-D Quantum Field Theory (QFT) generalizing the kinetic energy operators to an infinite series of higher derivatives inspired by string field theory and ghost-free non-local approaches to quantum gravity. At the perturbative level, the degrees of freedom of non-local Higgs are the same of the local theory. We prove that, at the non-perturbative level, the physical spectrum of the strongly-coupled Higgs mass is actually corrected from the "infinite number of derivatives" present in the action. The technique we use permits to derive the set of Dyson-Schwinger equations in differential form. This proves essentially useful when exact solutions to the local equations are known. We show that all the formalism of the local theory involving the Dyson-Schwinger approach extends quite naturally to the non-local case. Using these methods, the spectrum of the strongly-coupled non-local theories become accessible in the non-perturbative regimes. We calculate the N-point correlation functions and predict the mass-gap in the spectrum arising  purely from the self-interaction and the non-local scale M. The proper weak perturbation limit is correctly recovered from the strong coupling regime in the limit of the coupling of the theory goes to zero. Our results show the mass gap generated in the non-local theories gets damped in the UV and it reaches conformal limit. We discuss the implications of our result in particle physics and cosmology.
}

\end{abstract}

\maketitle

\section{Introduction}







A simple \textit{2-derivative} field theory works very well to explain local interactions in the classical and the quantum regimes. However,
the issue of re-normalization often encounters ultraviolet (UV) divergences, just like in the case of quantum electrodynamics. Particularly we find that in the context to the Standard Model (SM) of particle physics, this leads to the very infamous hierarchy
problem \cite{Haber:1984rc}. Solutions to this problem can be achieved via invoking several new particles at the or near the ElectroWeak (EW) scale, as for the case of supersymmetry (SUSY) etc.. However due to recent null new physics results from the Large Hadron Collider (LHC), alternative solutions are actively being pursued.

Either at the local level or at the global level, one may always go to beyond the \textit{2-derivative} kinetic term paradigm
which leads to certain classical and quantum instabilities depending on  the sign of the kinetic energy terms being negative or positive; \textit{for example}, the Ostrogradsky instability \cite{Ostrogradsky:1850fid} arising due to the energy density being unbounded from below at the classical level; this manifests as \textit{ghosts} at the quantum level \footnote{We call ghosts when there are extra propagating degrees of freedom, coming with a negative residue in the propagator.}. Very recently, higher-derivative approaches to such UV-completion of QFT have become popular. Initially, they were proposed as possible UV-regularized theoretical set-ups \cite{Moffat:1990jj, Evens:1990wf, Tomboulis:1997gg, Moffat:2011an, Tomboulis:2015gfa,Kleppe:1991rv} where infinite derivatives were motivated from p-adic string field theory \cite{sft1,sft2,sft3,padic1,padic2,padic3,Frampton-padic,Tseytlin:1995uq,marc,Siegel:2003vt,Calcagni:2013eua,Modesto:2011kw,Modesto:2012ga,Modesto:2015foa,Modesto:2017hzl}. These attempted diligently to address this divergence problem by generalizing the kinetic energy operators of the Standard Model (SM) to an infinite series of higher order derivatives suppressed by the scale of non-locality ($M$) at which the higher order derivatives come into play \cite{Biswas:2014yia}. Furthermore, one of the authors studied the negative running of the self-interacting term for the SM Higgs, which gives to rise to a metastable vacuum \cite{Olive:2016xmw}, was also cured by investigating the RGE of the theory \cite{Ghoshal:2017egr,Ghoshal:2020lfd}. It was discovered that the $\beta$-functions, at the scale of non-locality, and the wave-function re-normalization meant that the fields are frozen beyond $M$, and the theory reaches a conformal limit resolving the issue of Landau-poles \cite{Ghoshal:2020lfd}. Frankly speaking, capturing the infinite derivatives by exponential of an entire function softened UV behaviour in the desirable manner, without introducing any new degrees of freedom in the particle spectrum, as they contain no new poles in the propagators. These QFTs have been explicitly shown to be ghost-free \cite{Buoninfante:2018mre} and provides unique scattering phenomenology rendering transmutation of energy scales \footnote{See Ref. \cite{Salvio:2020axm} for a discussion on transmutation on energy scales in gravity and cosmology.} which has its own astrophysical \& cosmological phenomenology \cite{Buoninfante:2018gce} by one of the authors. Ref.\cite{Ghoshal:2018gpq}
showed the scale of non-locality maybe probed via dark matter (DM) detection experiments implications also studied by one of the authors.

In context to theories of gravity with such infinite derivative approch, Ref.\cite{Biswas:2011ar} investigated and was able to show that the most general quadratic curvature gravitational action (parity-invariant and torsion-free), with infinite series of covariant derivatives, can make the gravitational sector free from the Weyl {\it ghost} and it is also free from classical singularities, such as black hole singularities~\cite{Biswas:2011ar,Biswas:2013cha,Frolov:2015bia,Frolov:2015usa,Koshelev:2018hpt,Koshelev:2017bxd,Buoninfante:2018xiw,Cornell:2017irh,Buoninfante:2018rlq,Buoninfante:2018stt,Abel:2019zou,Buoninfante:2020ctr}~\footnote{In some previous studies, arguments were provided regarding non-singular solutions in Refs.~\cite{Tseytlin:1995uq,Siegel:2003vt}.} and cosmological singularities~\cite{Biswas:2005qr,Biswas:2006bs,Biswas:2010zk,Biswas:2012bp,Koshelev:2012qn,Koshelev:2018rau} \footnote{For implications in inflationary cosmology in context to infinite-derivative theories \& predictions in the CMB, see 
\cite{Koshelev:2016vhi,SravanKumar:2018dlo,Koshelev:2020fok,Koshelev:2020xby,Koshelev:2020foq}}. Recently, an astrophysical application was devised implying possible observational effects also in view of an explanation for dark matter \cite{Buoninfante:2019swn}. An extended group of symmetry, implying Galilean shift, was also proposed to obtain non-local extension of gravity theories \cite{Buoninfante:2018lnh}.

Naively, having infinite series of higher-order derivatives leads to having to solve the equations of motion with infinite initial values, but
on a general ground, the initial value problem for non-local field theories was studied in \cite{Barnaby:2007ve} where it was concluded that infinite derivative equations needs an finite number of initial data and this was illustrated further in context to non-local gravity shown in Ref. \cite{Calcagni:2018lyd}.  It has been shown, quite recently, that the unitarity issues are well addressed in Euclidean space and using Cutkosky rules, the results can be analytically continued to the Minkowski space \cite{Briscese:2018oyx,Briscese:2021mob,Koshelev:2021orf}. To be consistent with such results, we assume an Euclidean metric everywhere in this work.




The non-perturbative regime of non-local theories remains unexplored with only some studies attempting to address the question of the re-normalizability of some models of quantum gravity \cite{Calcagni:2018gke, Koshelev:2017tvv, Biswas:2006bs,Giacchini:2018wlf,Burzilla:2020utr} \footnote{4-derivative gravity theories have also recently been popularised as an alternatives approach to quantum gravity, but the theory consists of ghosts \cite{Salvio:2014soa,Salvio:2017qkx,Salvio:2018crh,Anselmi:2018tmf,Anselmi:2018ibi}.}. In this paper we provide for the first time an approach to compute the physical spectrum induced in the non-perturbative regimes of the non-local field theory. This approach relies on studies, performed in the last two decades, about local non-linear quantum field theories \cite{Frasca:2019ysi, Chaichian:2018cyv, Frasca:2017slg, Frasca:2016sky, Frasca:2015yva, Frasca:2015wva, Frasca:2013tma, Frasca:2012ne, Frasca:2009bc, Frasca:2010ce, Frasca:2008tg, Frasca:2009yp, Frasca:2008zp, Frasca:2007uz, Frasca:2006yx, Frasca:2005sx, Frasca:2005mv, Frasca:2005fs}. The main technique we will rely on is the Bender, Milton, Savage method for Dyson-Schwinger equations as devised in \cite{Bender:1999ek} that is widely used in the aforementioned studies. We will see that this technique extends quite naturally to non-local theories. Also, we will describe a perturbation approach holding in the formal limit of the coupling going to infinity, as widely discussed in the references already mentioned, but applied to the non-local case.

Our approach starts from exact solutions of the local theory as devised in \cite{Frasca:2015yva}. This means that the only limitation we are considering arises from adopting the condition that non-local effect are so small to make our perturbation approach meaningful. Then, we are working completely in Euclidean metric where such non-local theories are meaningful and, once we conclude our analysis a analytical continuation is taken to move to the Minkowski space as done in most of the studies about these theories in the current literature. Therefore, as non-local effects are taken to be really small as not yet seen experimentally, we can assume that the propagator of the theory is just lightly deformed by such effects and so, an interpretation of the poles of the Green function as excitations of the theory could be supported yet. In this sense, a mass gap for the non-local theory has the same meaning as for the local case.
Considering non-local effect as small and slightly deforming the local theory implies that the boundary problem for Dyson-Schwinger equations can be safely ignored at this stage. 

The paper is organized as follows: we describe scalar infinite derivative model in section \ref{Review}. In section \ref{Classical1}, we give an overview of technique that applies in non-perturbative regimes of local field theories. In section \ref{Classical2}, we apply the techniques discussed in the preceding section to non-local infinite derivative scalar theory at the classical level. In section \ref{Quantum}, we extend our analysis to the quantum theory by computing the corresponding correlation functions till 2-point. In the final section, we conclude by discussing some of the impacting aspects of our study in particle physics and cosmology.

\medskip

\section{Infinite Derivative Higgs: Review}
\label{Review}

Let us start with the action for the infinite derivative theory given by \cite{Biswas:2014yia}:
\begin{equation} \label{Action}
    S = \int d^4 x\ \left(-\frac{1}{2} \phi e^{f(\Box)}(\Box+m^2)\phi -\frac{\lambda}{4!}\phi^{4}\right)
\end{equation} 
The normalization of $\phi$ is so chosen in such a manner that the residue at the $p^2=m^2$ pole is unity in this case.
$\Box= \eta_{\mu\nu}\partial^{\mu}\partial^{\nu}$ $(\mu, \nu=0,1,2,3$) with the convention of the metric signature $(+,-,-,-)$, 
  $m$ is the mass of the scalar particle, 
  and $M$ is the energy scale of the non-locality which is taken to be below the Planck scale. 
As you may note n the above equation, the kinetic energy terms are generalized with higher derivatives suppressed by the non-local energy scale $M$, whilst the scalar self-interaction is the typical quartic potential one. 
This action is reduced into the standard local field theory in the limit of $M \to \infty$.
Regarding the non local form-factor $e^{f(\Box)}$, various proposals have been considered in the literature \cite{Edholm:2016hbt}. For simplicity, we will use  $e^{-\frac{\Box^2}{M^2}}$. But overall, what is essential is that the factor is an entire function without zeros avoiding in this way possible ghosts in the spectrum of the theory.

In Euclidean space ($p^0\rightarrow ip_{E} ^0$) the propagator is given by,
\be \label{Prop}
\Pi(p^2)=-\frac{ie^{f(-p_E^2)}}{p_E^2+m^2}
\ee
while the vertex factor is, as usual, given by $-i\lambda$.
Note that the non-local extensions of the local theory leads to the exponential suppression of the propagators for $p_E^2 > M^2$ region, 
and this fact means that the quantum corrections in the UV will be frozen at energies higher than $M$ (basically all the $\beta-$functions vanish beyond the scale of non-locality M making the theory reach an asymptotically conformal limit in the UV) \cite{Ghoshal:2017egr,Ghoshal:2020lfd}. 

Although the action in Eqn.(\ref{Action}) shows a  modification in the kinetic term, 
however, note that an equivalent description can be done with the usual local Klein-Gordon kinetic operator via the following field re-definition \cite{Buoninfante:2018mre}:
\begin{equation}
\begin{array}{rl}
\tilde{\phi}(x)= & \displaystyle e^{-\frac{1}{2}f(\Box)}\phi(x)\\
= & \displaystyle \int d^4y \mathcal{F}(x-y)\phi(y),
\end{array}
\label{42}
\end{equation}
where $\mathcal{F}(x-y):=e^{-\frac{1}{2}f(\Box)}\delta^{(4)}(x-y)$. 
However, the real fields of the theory are $\phi$, and not $\tilde{\phi}$. With the field redefinition into the action it becomes:
\begin{equation}
S=-\frac{1}{2}\int d^4x \left(\tilde{\phi}(x)(\Box+m^2)\tilde{\phi}(x) -\int d^4x \frac{ \lambda }{4!}\left(e^{\frac{1}{2}f(\Box)}\tilde{\phi}(x)\right)^4
+j(x)e^{\frac{1}{2}f(\Box)}\tilde{\phi}(x)\right)
\label{43}
\end{equation}
%
where we added a source term $j(x)$ \footnote{We will work in the Euclidean space, \& in defining the relation between Minkowski and Euclidean metrics, we will assume the conclusions given in \cite{Pius:2016jsl,Briscese:2018oyx} which is done via analytical continuation. Technically, at least in our computations, it is fine to go with a standard Wick rotation.}.

Eqn.\eqref{43} shows that the form factor $e^{\frac{1}{2} f(\Box)}$ appears in the interaction term 
, thereby non-locality is significant only when 
the interaction is switched on as the free-part remains just the standard local Klein-Gordon kinetic term. 

We use the former description using the $\phi$ field definition throughout the paper. 

\medskip

\section{Techniques in strongly coupled local field theories}
\label{Classical1}

Following Refs. \cite{Frasca:2015yva,Frasca:2015wva,Frasca:2013tma}, we present here some of the results obtained in the local theory that we will use in the rest of the paper for computations in the non-local theory.
We consider the following action
\be
S=\int d^4x\left(\frac{1}{2}(\partial\phi)^2-\frac{\lambda}{4}\phi^4+j\phi\right).
\ee
Therefore, the scalar theory we consider has the following equation of motion
\be
\label{eq:phi3}
-\Box\phi(x)=-\lambda\phi^3(x)+j(x).
\ee
We would like to do perturbation theory in the formal limit $\lambda\rightarrow\infty$. For our aims, we rescale $x\rightarrow\sqrt{\lambda}x$ and we put
\be
\phi(x)=\sum_{k=0}^\infty\lambda^{-k}\phi_k(x).
\ee
A direct substitution will yield the set of equations
\bea
-\Box\phi_0(x)&=&-\phi_0^3(x) \nonumber \\
-\Box\phi_1(x)&=&-3\phi_0^2(x)\phi_1(x)+j(x) \nonumber \\
-\Box\phi_2(x)&=&-3\phi_0^2(x)\phi_2(x)-3\phi_0(x)\phi_1^2(x) \nonumber \\
&\vdots&.
\eea
We see here how we fixed the ordering for the source $j(x)$. This is a lucky case because we know both the solutions to the leading order equation and for the 2-point function in the following equations \cite{Frasca:2009bc} and so, we can have a formal solution to the full set of pertubative equations.

It is important to point out that identical equations for $\phi_n(x)$ can be obtained if we consider Eqn.(\ref{eq:phi3}) as a functional equation with $\phi=\phi[j]$ and take the following Taylor series \cite{Frasca:2013tma}
\be
\phi[j]=\phi[0]+\int d^4y\left.\frac{\delta\phi}{\delta j(y)}\right|_{j=0}j(y)
+\frac{1}{2}\int d^4yd^4z\left.\frac{\delta^2\phi}{\delta j(y)\delta j(z)}\right|_{j=0}j(y)j(z)
+O(j^3).
\ee
We have immediately the definition of all the correlation functions for the classical theory. Specially, for the 2-point function is
\be
G_c^{(2)}(x-y)=\left.\frac{\delta\phi}{\delta j(y)}\right|_{j=0},
\ee
and $\phi[0]=\phi_0(x)$. To avoid confusion, we will rename $\phi_0=\phi_c$ in the following. 
The mass gap arises naturally, due to the non-linear self-interaction term, as
\be
\label{eq:phic}
\phi_c(x)=\mu\left(\frac{2}{\lambda}\right)^\frac{1}{4}{\rm sn}(p\cdot x+\theta,i)
\ee
where sn is a Jacobi elliptical function, $\mu$ and $\theta$ two arbitrary integration constants.  This holds provided the following relation is obeyed \cite{Frasca:2009bc}:
\be
\label{eq:DS}
p^2=\mu^2\sqrt{\frac{\lambda}{2}}.
\ee
What we have got are massive non-linear waves extending to all the space-time. The analog of a free theory are plane waves. It is important to emphasize that $\mu$ fixes the energy scale and $\theta$ grants the translation invariance of the theory in the higher order equations.

The 2-point function can be easily obtained as its equation has an exact solution. We write it down in the momenta space \cite{Frasca:2009bc}:
\begin{equation}
\label{eq:G1c}
    G_c^{(1)}(k)=\sum_{n=0}^\infty\frac{B_n}{k^2-m_n^2+i\epsilon}
\end{equation}
with
\begin{equation}
    B_n=(2n+1)^2\frac{\pi^2}{2K^2(i)}\frac{e^{-\left(n+\frac{1}{2}\right)\pi}}{1+e^{-(2n+1)\pi}}
\end{equation}
and
\begin{equation}
    m_n=(2n+1)\frac{\pi}{2K(i)}\left(\frac{\lambda}{2}\right)^\frac{1}{4}\mu.
\end{equation}
We can see that, in the limit $\mu\rightarrow 0$ the free theory is recovered at the leading order for the weak perturbation limit $\lambda\rightarrow 0$. Anyhow, in such a case the exact solution has the simple form $\phi(x)\sim constant+\mu t+O(\lambda)$ holding for $\lambda\rightarrow 0$. The propagator then will be given just by the massless case $\sim 1/k^2$ since $\sum_{n=0}^\infty B_n=1$. So, in the local theory case, we see that we correctly recover the proper weak field limit. This will continue to be true in the non-local scenario as well since we will use this exact solution
as an input.

For the quantum case, we have given the results and a description of the Bender-Milton-Savage method into Appendix A. The first two Dyson-Schwinger equations take the form
\be
\partial^2G_1(x)+\lambda[G_1(x)]^3+3\lambda G_2(0)G_1(x)+G_3(0,0)=0.
\ee
and
\be
\partial^2G_2(x,y)+3\lambda[G_1(x)]^2G_2(x,y)+
3\lambda G_3(0,y)G_1(x)
+3\lambda G_2(0)G_2(x,y)
+G_4(0,0,y)=
\delta^4(x-y).
\ee
These two equations are identical to the classical ones except for a re-normalization effect on the mass of the field that is shifted. This yields a gap equation for the theory that can be written down as \cite{Frasca:2017slg}
\bea
&\delta m^2=3\lambda\sqrt{\delta m^2+\mu^2\sqrt{\lambda/2}}Z(\delta m^2,\lambda)\times &\\
&\int\frac{d^4p}{(2\pi)^4}\sum_{n=0}^\infty (-1)^n(2n+1)^2\frac{2\pi^3}{K^3(k^2(\delta m))}\times \nonumber &\\
&\frac{q^{n+1/2}}{1-q^{2n+1}}\frac{1}{p^2-m_n^2+i\epsilon}. \nonumber
\eea
where $K(k)$ is the complete integral elliptic of the first kind, $Z(\delta m^2,\lambda)$ a constant multiplying the propagator equation, and



\begin{equation} 
k^2(\delta m)=\frac{\delta m^2-\sqrt{\delta m^4+2\lambda\mu^4}}{\delta m^2+\sqrt{\delta m^4+2\lambda\mu^4}},
\end{equation}
and 
\begin{equation}
\lim_{\delta m\rightarrow 0}\sqrt{\delta m^2+\mu^2\sqrt{\lambda/2}}Z(\delta m^2,\lambda)=\frac{1}{8},
\end{equation}
having set $q=\exp\left(-\pi K'(k^2)/K(k^2)\right)$ with $K'(k^2)=K(1+k^2)$. 
These are the key equations to compute $\delta m$ assuming it is small. \footnote{It was recently shown that the agreement with lattice computations for Yang-Mills theory, with this equation properly re-scaled and derived from quantum theory of Yang-Mills equations, is astonishingly good \cite{Frasca:2017slg,Frasca:2016sky}.}


In the following section we move on to the case for the non-local theory.

\medskip

\section{Mass Gap in Non-Local Theory: Classical Case}
\label{Classical2}


The action in Eqn.(\ref{Action}) leads to the equation of motion for $m=0$: \footnote{For our convenience, we remove the $1/2$ factor in the exponential, from here-on. This amounts to an inessential re-scaling of the function f.}
\begin{equation}
\label{eq:mot}
    -\Box\phi=-\frac{\lambda}{3!}e^{f(\Box)}\left(e^{f(\Box)}\phi \right)^3+e^{f(\Box)}j
\end{equation}
where we have added an arbitrary source \textit{j} to be set to zero to the end of computations. We have chosen a local current for our aims as we consider non-local effects at sufficient high-energies to neglect non-local effects on the sources. Our aim is to consider the field $\phi$ as a functional $\phi[j]$ and to study a functional Taylor series on $j$.

The main point is to consider a background field approach where we develop a series starting from a non-perturbative solution of the local case. This can be accomplished by re-writing the above equation as
\begin{equation}
\label{eq:phi_start}
    -\Box\phi+\frac{\lambda}{3!}\phi^3=-\frac{\lambda}{3!}e^{f(\Box) }\left(e^{f(\Box)}\phi \right)^3+\frac{\lambda}{3!}\phi^3+e^{f(\Box)}j.
\end{equation}
The idea is that non-local effects are small and then, our solution should not be too far from that of the local case that we know. This is the reason why we introduce the term $\frac{\lambda}{3!}\phi^3$ and consider all the rest as a perturbation. Therefore, we will work out our perturbation series starting from the solutions of the equation
\begin{equation}
    -\Box\phi_c+\frac{\lambda}{3!}\phi_c^3=0.
\end{equation}
This solution, obtained by properly re-scaling the coupling in Eqn.(\ref{eq:phic}), can be written down as a Fourier series
\begin{equation}
\label{eq:sns}
    \phi_c(x)=\mu\left(\frac{1}{3\lambda}\right)^\frac{1}{4}\frac{2\pi}{K(i)}\sum_{n=0}^\infty(-1)^n\frac{e^{-\left(n+\frac{1}{2}\right)\pi}}{1+e^{-(2n+1)\pi}}
    \sin\left((2n+1)\frac{\pi}{2K(i)}(p\cdot x+\theta)\right),
\end{equation}
being $K(i)$ the complete elliptic integral of the first kind. It should be noted that this series is absolutely convergent.
The equation for the 2-point function can be obtained by the functional derivative with respect to $j(y)$ of Eqn.(\ref{eq:phi_start}).




\subsubsection{Classical Strong Perturbation Series}


We follow the methods as in the usual perturbation theory in the formal limit $\lambda\rightarrow\infty$  \cite{Frasca:2013tma}. 
The technique is same as already described in the preceding section for the local case but, here, we have considered the effects of the non-locality that are evaluated as small deviations from the local limit itself. Our results become the same as the local theory in the limit $M \rightarrow \infty$.

This can be obtained after a re-scaling $x\rightarrow\sqrt{\lambda}x$ and taking
\begin{equation}
    \phi(x)=\sum_{n=0}^\infty\lambda^{-n}\phi_n(x).
\end{equation}
This yields the following set of perturbation equations (we have retained the coupling $\lambda$ for future convenience but the ordering should be clear)
\begin{eqnarray}
\label{eq:phi0phi1}
-\Box\phi_0&=&-\frac{\lambda}{3!}e^{f(\Box)}\left(e^{f(\Box)}\phi_0\right)^3 \nonumber \\
-\Box\phi_1&=&-\frac{\lambda}{2!}e^{f(\Box)}\left[\left(e^{f(\Box)}\phi_0\right)^2e^{f(\Box)}\phi_1\right]+e^{f(\Box)}j(x) \nonumber \\
&\vdots&. 
\nonumber \\
-\Box\phi_n&=&-\frac{\lambda}{3!}e^{f(\Box)}\left[\frac{1}{ne^{f(\Box)}\phi_0(x)}\sum_{k=1}^n(3k-n+k)e^{f(\Box)}\phi_k(x)\chi_{n-k}(x)\right] 
\end{eqnarray}
where
\begin{equation}
\chi_m(x)=\frac{1}{me^{f(\Box)}\phi_0(x)}\sum_{k=1}^m(kp-m+k)e^{f(\Box)}\phi_k(x)\chi_{m-k}(x)\qquad m\ge 1.
\end{equation}
where we have used the following recursion relations \cite{grad} \footnote{We set $x=1$ at the end of computations taking the convergence of these series for granted}:
\begin{equation}
\label{eq:sump}
    \left(\sum_{n=0}^\infty a_nx^n\right)^p=\sum_{n=0}^\infty c_nx^n
\end{equation}
being $c_0=a^p$ and given the recurrence relation
\begin{equation}
\label{eq:rec}
c_m=\frac{1}{ma_0}\sum_{k=1}^m(kp-m+k)a_kc_{m-k}\qquad m\ge 1.
\end{equation}
The same set of equations can be obtained as a functional Taylor series in $j$, having $\phi_n$ as coefficients \cite{Frasca:2013tma}.

\subsubsection{Leading Order Equation}

In order to understand the full physical spectrum and the mass gap in the theory, we solve the first equation in Eqn.(\ref{eq:phi0phi1}), re-writing as
\begin{equation}
    -\Box\phi_0+\frac{\lambda}{3!}\phi_0^3=-\frac{\lambda}{3!}e^{f(\Box)}\left(e^{f(\Box)}\phi_0\right)^3+\frac{\lambda}{3!}\phi_0^3
\end{equation}
and expanding as perturbative series leading to
\begin{eqnarray} \label{lo_eqn}
     -\Box\phi_c+\frac{\lambda}{3!}\phi_c^3&=&0 \nonumber \\
     -\Box\phi_0^{(1)}+\frac{\lambda}{2!}\phi_c^2\phi_0^{(1)}&=&-\frac{\lambda}{3!}e^{f(\Box)}\left(e^{f(\Box)}\phi_c\right)^3+\frac{\lambda}{3!}\phi_c^3 \nonumber \\
     &\vdots&
\end{eqnarray}
This set is consistent with our previous expansion wherever non-local effects can be considered as a small corrections to the local solutions.

Using the solution for $\phi_c$, as per Eqn.(\ref{eq:sns}), we will have, from the R.H.S. of the second equation in Eqn.(\ref{lo_eqn}), to be
\begin{eqnarray}
 e^{f(\Box)}\left(e^{f(\Box)}\phi_c\right)^3
 &=&\mu^3\left(\frac{1}{3\lambda}\right)^\frac{3}{4}\frac{8\pi^3}{K^3(i)}
 e^{f(\Box)}\left[\sum_{n=0}^\infty(-1)^n\frac{e^{-\left(n+\frac{1}{2}\right)\pi}}{1+e^{-(2n+1)\pi}}e^{f\left(-(2n+1)^2\frac{\pi^2}{4K^2(i)}p^2\right)}
 \times
 \right. \nonumber \\
 &&\left.\sin\left((2n+1)\frac{\pi}{2K(i)}(p\cdot x+\theta)\right)\right]^3 \nonumber \\
 &=& \mu^3
 \left(\frac{1}{3\lambda}\right)^\frac{3}{4}\frac{8\pi^3}{K^3(i)} \nonumber \\
&&\left[\frac{e^{-\frac{3\pi}{2}}}{(1+e^{-\pi})^3}e^{3f\left(-\frac{\pi^2}{4K^2(i)}p^2\right)}\sin^3\left(\frac{\pi}{2K(i)}(p\cdot x+\theta)\right)+
\right. \nonumber \\
&&\left.\sum_{n=1}^\infty C_n(x) ,
\right]
\end{eqnarray}
where use has been made again of Eqn.(\ref{eq:sump}-\ref{eq:rec}). We have interchanged the infinite-derivative factor and the infinite summation assuming this correct as the series, is absolutely convergent as stated before. 
Similarly, the coefficients $C_n$ can be computed using Eqn.(\ref{eq:rec}) yielding the recurrence relation
\begin{equation}
     C_{m}(x)=\frac{1}{ma_0}\sum_{k=1}^{m}(3k-m+k)a_k(x)c_{m-k}(x)
\end{equation}
Therefore we can derive at all the terms of the solution. E.g., the next term is simply given by
\begin{eqnarray}
    C_1(x)&=&3a_1a_0^2=\nonumber \\
    &&-3\frac{e^{-\pi}}{(1+e^{-\pi})^2}e^{2f\left(-\frac{\pi^2}{4K^2(i)}p^2\right)}}
    \frac{e^{-\frac{3}{2}\pi}}{1+e^{-3\pi}}e^{f\left(-9\frac{\pi^2}{4K^2(i)}p^2\right)\times \nonumber \\
   &&\sin^2\left(\frac{\pi}{2K(i)}(p\cdot x+\theta)\right)
    \sin\left(3\frac{\pi}{2K(i)}(p\cdot x+\theta)\right).
\end{eqnarray}
We see that such terms are heavily damped in an exponential manner. 

\smallskip

Furthermore, noting the identity,
\begin{equation}
    (\operatorname{sn}(x,i))''=-2\operatorname{sn}^3(x,i)
\end{equation}
one may straight-forwardly reach the field solutions
\begin{equation}
  \phi_c^3(x)=\mu^3\left(\frac{1}{3\lambda}\right)^\frac{3}{4}\frac{\pi^5}{K^5(i)}\sum_{n=0}^\infty(-1)^n(2n+1)^2\frac{e^{-\left(n+\frac{1}{2}\right)\pi}}{1+e^{-(2n+1)\pi}}
    \sin\left((2n+1)\frac{\pi}{2K(i)}(p\cdot x+\theta)\right).  
\end{equation}
This will give us the final equation to solve.


\subsubsection{Solution to the Final Equation}

The final equation that is to be solved becomes:

\begin{eqnarray}
    & & -\Box\phi_0^{(1)}+\frac{\lambda}{2!}\phi_c^2\phi_0^{(1)} = -
    \mu^3
 \left(\frac{\lambda}{27}\right)^\frac{1}{4}\frac{4\pi^3}{3K^3(i)}
 \left[\sum_{n=1}^\infty C_n(x)\right]+
\nonumber \\
&&\mu^3\left(\frac{\lambda}{27}\right)^\frac{1}{4}
    \left[1-\frac{4\pi^3}{3K^3(i)}\frac{e^{-\frac{3\pi}{2}}}{(1+e^{-\pi})^3}e^{3f\left(-\frac{\pi^2}{4K^2(i)}p^2\right)}\right]\sin^3\left(\frac{\pi}{2K(i)}(p\cdot x+\theta)\right). \nonumber
\end{eqnarray}
It is interesting to note the re-normalization term to the classical potential arising from the higher derivatives in the theory.
This can be evaluated on-shell using the dispersion relation technique (See Eqn.\ref{eq:DS}).

\subsubsection{Classical 2-point-function Solution}

Starting from Eqn.(\ref{eq:mot}), we take the functional derivative with respect to $j$. This yields\footnote{It should be pointed out that, in the current literature on non-local infinite-derivative field theories, there is a somewhat cavalier treatment of distributions. Indeed, it is a well-acquired result that the propagator equation has a delta function multiplied by the entire function yielding an exponential factor in the momenta space \cite{Moffat:1990jj, Evens:1990wf, Tomboulis:1997gg, Moffat:2011an, Tomboulis:2015gfa,sft1,sft2,sft3,padic1,padic2,padic3,Frampton-padic,Tseytlin:1995uq,marc,Siegel:2003vt}. Often, this is neither explicitly discussed nor mathematically deepened and should be considered as an open problem. }
\begin{equation}
\label{eq:G}
    -\Box G=-\frac{\lambda}{2!}e^{f(\Box)}\left[\left(e^{f(\Box)}\phi_0\right)^2e^{f(\Box)}G\right]+e^{f(\Box)}\delta^4(x-y).
\end{equation}
which is the same obtained perturbatively in Eqn.(\ref{eq:phi0phi1}). The same equations one has the leading order solution
\begin{equation}
    -\Box\phi_0=-\frac{\lambda}{3!}e^{f(\Box)}\left(e^{f(\Box)}\phi_0\right)^3.
\end{equation}
We have already treated this equation by an iterative method starting from exact classical local solutions. We are assuming, as in Ref.\cite{Buoninfante:2018mre}, that non-local effects should represents corrections to the known local solutions. 

The equation for the classical 2-point function is linear. We could apply a Fourier transform straightforwardly 
\cite{Avramidi:2000bm}. Firstly, we note that, from the leading order solution is
\begin{equation}
    \phi_0(x)=\phi_c(x)+\chi
\end{equation}
where the correction $\chi$ due to the infinite derivatives can be written as
\begin{eqnarray}
    \chi &=& -\mu^3
 \left(\frac{\lambda}{27}\right)^\frac{1}{4}\frac{4\pi^3}{3K^3(i)}\times \\
 &&\int d^4y\Delta(x-y)\left\{\sum_{n=1}^\infty C_n(y)\right.+
\nonumber \\
&&\left.\mu^3\left(\frac{\lambda}{27}\right)^\frac{1}{4}
    \left[1-\frac{4\pi^3}{3K^3(i)}\frac{e^{-\frac{3\pi}{2}}}{(1+e^{-\pi})^3}e^{3f\left(-\frac{\pi^2}{4K^2(i)}p^2\right)}\right]\sin^3\left(\frac{\pi}{2K(i)}(p\cdot y+\theta)\right)\right\}. \nonumber
\end{eqnarray}
Here $\Delta(x-y)=G_c^{(1)}(x-y)$ as given in eq.(\ref{eq:G1c}).


Then, we can evaluate the rhs of eq.(\ref{eq:G}) as
\begin{eqnarray}
    e^{f(\Box)}\left[\left(e^{f(\Box)}\phi_0\right)^2e^{f(\Box)}G\right]
    =
    e^{f(\Box)}\left[\left(e^{f(\Box)}\phi_c+e^{f(\Box)}\chi\right)^2e^{f(\Box)}G\right]\approx&&\nonumber \\
    e^{f(\Box)}\left[\left(e^{f(\Box)}\phi_c\right)^2e^{f(\Box)}G   \right]+2e^{f(\Box)}\left[\left[(e^{f(\Box)}\phi_c)\left(e^{f(\Box)}\chi\right)\right]e^{f(\Box)}G\right].&&
\end{eqnarray}
This means that we are considering $\chi$ as a higher order correction according to our previous computation in perturbation theory. This implies we can consider the notion that non-local effects will arise as a correction to the local theory. Nonetheless, we notice that, already at the leading order, one gets corrections to the spectrum of the theory. Then, the equation for the 2-point function can be written as
\begin{equation}
    -\Box G=-\frac{\lambda}{2!}e^{f(\Box)}\left[\left(e^{f(\Box)}\phi_c\right)^2e^{f(\Box)}G\right]-\lambda e^{f(\Box)}\left[(e^{f(\Box)}\phi_c)\left(e^{f(\Box)}\chi\right)e^{f(\Box)}G\right]+O(\chi^2)+e^{f(\Box)}\delta^4(x-y).  
\end{equation}
Then, we evaluate
\begin{eqnarray}
    \left(e^{f(\Box)}\phi_c\right)^2&=&
   \nonumber \\
     \left[
    \mu\left(\frac{1}{3\lambda}\right)^\frac{1}{4}\frac{2\pi}{K(i)}\sum_{n=0}^\infty(-1)^n\frac{e^{-\left(n+\frac{1}{2}\right)\pi}}{1+e^{-(2n+1)\pi}}
    e^{f\left(-(2n+1)\frac{\pi}{2K(i)}p^2\right)}\right.\times&& \nonumber \\
    \left.\sin\left((2n+1)\frac{\pi}{2K(i)}(p\cdot x+\theta)\right)\right]^2.&&
\end{eqnarray}
At this stage, we have from Eqn.(\ref{eq:sump})
\begin{eqnarray}
 \left(e^{f(\Box)}\phi_c\right)^2&=&\mu^2
 \left(\frac{1}{3\lambda}\right)^\frac{1}{2}\frac{4\pi^2}{K^2(i)}\times \nonumber \\
&&\left[\frac{e^{-\pi}}{(1+e^{-\pi})^2}e^{2f\left(-\frac{\pi^2}{4K^2(i)}p^2\right)}\sin^2\left(\frac{\pi}{2K(i)}(p\cdot x+\theta)\right)+
\right. \nonumber \\
&&\left.\sum_{n=1}^\infty D_n(x)
\right].
\end{eqnarray}
with the coefficients $D_{n}$ given by eq.(\ref{eq:rec}). This yields for the 2-point function a forced Mathieu-like equation as
\begin{eqnarray}
    &&-\Box G+\frac{\lambda^\frac{1}{2}}{2\sqrt{3}}\frac{\pi^2}{K^2(i)}
    \frac{e^{-\pi}}{(1+e^{-\pi})^2}e^{2f\left(-\frac{\pi^2}{4K^2(i)}p^2\right)}\times \nonumber \\
     &&e^{f(\Box)}\left\{\left[1-\cos\left(\frac{\pi}{K(i)}(p\cdot x+\theta)\right)\right]e^{f(\Box)}G\right\}= \nonumber \\
    &&-\frac{\lambda^\frac{1}{2}}{2\sqrt{3}}\sum_{n=1}^\infty D_n(x)
 +O(\chi)+e^{f(\Box)}\delta^4(x-y).
\end{eqnarray}
From this we can get a set of eigenvalues that apply to the non-local classical theory. We will solve it using a Fourier transform.

By using translation invariance we fix the phase $\theta$ so to take consistently $y=0$. Then, we Fourier transform to get
\begin{eqnarray}
    k^2G(k)+m_0^2e^{2f(-k^2)}G(k)-\frac{m_0^2}{2}e^{f(-k^2)}\left[e^{f(-(k-\eta p)^2)}G(k-\eta p)+e^{f(-(k+\eta p)^2)}G(k+\eta p)\right]&=& \nonumber \\
    e^{f(-k^2)}+{\cal F}\left(h.o.c.\right).&&
\end{eqnarray}
Here $h.o.c.$ stands for higher order corrections and $\eta=\pi/K(i)$. We notice immediately that the theory developed a mass gap given by
\begin{equation}
\label{eq:m0}
    m_0^2=\frac{\lambda^\frac{1}{2}}{2\sqrt{3}}\frac{\pi^2}{K^2(i)}
    \frac{e^{-\pi}}{(1+e^{-\pi})^2}e^{3f\left(-\frac{\pi^2}{4K^2(i)}p^2\right)}\mu^2.
\end{equation}
Anyways, we can write the equation for the propagator as
\begin{eqnarray}
    G(k)&=&\frac{e^{f(-k^2)}}{k^2+m_0^2e^{2f(-k^2)}}\times \\ &&\left[1+\frac{m_0^2}{2}e^{f(-k^2)}\left(e^{f(-(k-\eta p)^2)}G(k-\eta p)+e^{f(-(k+\eta p)^2)}G(k+\eta p)\right)\right]+{\cal F}\left(h.o.c.\right). \nonumber
\end{eqnarray}
Let us consider the leading order equation
\begin{equation}
\label{eq:G0}
    G_0(k)=\Delta(k)\left[1+\frac{m_0^2}{2}e^{f(-k^2)}\left(e^{f(-(k-\eta p)^2)}G_0(k-\eta p)+e^{f(-(k+\eta p)^2)}G_0(k+\eta p)\right)\right].
\end{equation}
being
\begin{equation}
\label{eq:G_lo}
    \Delta(k)=\frac{e^{f(-k^2)}}{k^2+m_0^2e^{2f(-k^2)}}.
\end{equation}


Eq.(\ref{eq:G0}) can be solved using an iterative procedure. This is obtained by computing
\begin{equation}
    G_0(k-\eta p)=\Delta(k-\eta p)\left[1+\frac{m_0^2}{2}e^{f(-(k-\eta p)^2)}\left(e^{f(-(k-2\eta p)^2)}G_0(k-2\eta p)+e^{f(-k^2)}G_0(k)\right)\right]
\end{equation}
and
\begin{equation}
    G_0(k+p)=\Delta(k+\eta p)\left[1+\frac{m_0^2}{2}e^{f(-(k+\eta p)^2)}\left(e^{f(-k^2)}G_0(k)+e^{f(-(k+2\eta p)^2)}G_0(k+2\eta p)\right)\right].
\end{equation}
Then, we put this into eq.(\ref{eq:G0}) yielding
\begin{eqnarray}
    G_0(k)&=&\Delta(k)\times \\
    &&\left[
    1+\frac{m_0^2}{2}e^{f(-k^2)}\times\right. \nonumber \\
    &&\left(
    \Delta(k-2\eta p)e^{f(-(k-2\eta p)^2)}
    \left[
    1+\frac{m_0^2}{2}e^{f(-(k-\eta p)^2)}\times\right.\right. \nonumber \\
    &&\left.\left(
    e^{f(-(k-2\eta p)^2)}G_0(k-2\eta p)+e^{f(-k^2)}G_0(k)
    \right)
    \right]
    + \nonumber \\
    &&\Delta(k+\eta p)e^{f(-(k+\eta p)^2)}
    \left[
    1+\frac{m_0^2}{2}e^{f(-(k+\eta p)^2)}\times\right. \nonumber \\
    &&\left.\left.\left.\left(
    e^{f(-k^2)}G_0(k)+e^{f(-(k+2\eta p)^2)}G_0(k+2\eta p)
    \right)
    \right]
    \right)
    \right]. \nonumber
\end{eqnarray}
This yields
\begin{eqnarray}
    G_0(k)
    \left[
    1-\frac{m_0^4}{4}e^{2f(-k^2)}\Delta(k)
    \left(
    \Delta(k+\eta p)
    e^{2f(-(k+\eta p)^2)}+
    \Delta(k-\eta p)e^{2f(-(k-\eta p)^2)}
    \right)
    \right]
    &=&
    \Delta(k) \nonumber \\
    \left[
    1+\frac{m_0^2}{2}e^{f(-k^2)}
    \left(
    e^{f(-(k-\eta p)^2)}\Delta(k-\eta p)
    \left[
    1+\frac{m_0^2}{2}e^{f(-(k-2p)^2)}G_0(k-2p)
    \right]
    \right.
    \right.
    +&& \nonumber \\
    \left.\left.
    e^{f(-(k+\eta p)^2)}\Delta(k+\eta p)
    \left[
    1+\frac{m_0^2}{2}e^{f(-(k+2p)^2)}G_0(k+2p)
    \right]
    \right)
    \right].&&
\end{eqnarray}
This will give in the end
\begin{eqnarray}
\label{eq:G0R}
    G_0(k)&=&
    \frac{\Delta(k)}{
    1-\frac{m_0^4}{4}e^{2f(-k^2)}\Delta(k)
    \left(
    \Delta(k+\eta p)
    e^{2f(-(k+\eta p)^2)}+
    \Delta(k-\eta p)e^{2f(-(k-\eta p)^2)}
    \right)}
    \times \nonumber \\
    &&\left[
    1+\frac{m_0^2}{2}e^{f(-k^2)}
    \left(
    e^{f(-(k-\eta p)^2)}\Delta(k-\eta p)
    \left[
    1+\frac{m_0^2}{2}e^{f(-(k-2\eta p)^2)}G_0(k-2\eta p)
    \right]
    \right.
    \right.
    + \nonumber \\
    &&\left.\left.
    e^{f(-(k+\eta p)^2)}\Delta(k+\eta p)
    \left[
    1+\frac{m_0^2}{2}e^{f(-(k+2\eta p)^2)}G_0(k+2\eta p)
    \right]
    \right)
    \right].
\end{eqnarray}
The leading term is given by
\begin{equation} \label{result_classicl}
    \Delta_R(k)=
    \frac{e^{f(-k^2)}}{k^2+m_0^2e^{2f(-k^2)}-\frac{m_0^4}{4}e^{2f(-k^2)}
    \left(
    \Delta(k+\eta p)e^{2f(-(k+\eta p)^2)}+\Delta(k-\eta p)e^{2f(-(k-\eta p)^2)}
    \right)}.
\end{equation}
The mass gets renormalized by a finite term. Poles of this propagator yield higher order excitations. Although this procedure is somewhat cumbersome, in principle, it can be iterated as many times as one wants to obtain the exact propagator of the theory and the corresponding spectrum. This propagator, at this iteration stage, can be written down in the form
\begin{equation} \label{final_classical}
    \Delta_R(k)=
    \frac{e^{f(-k^2)}}{k^2+m_0^2e^{2f(-k^2)}}\frac{1}{1-\Pi(k)},
\end{equation}
where
\be
  \Pi(k)=\Delta(k)\frac{m_0^4}{4}e^{2f(-k^2)}
    \left(
    \Delta(k+\eta p)e^{2f(-(k+\eta p)^2)}+\Delta(k-\eta p)e^{2f(-(k-\eta p)^2)}
    \right).
\ee
We will recognize here a pole and we are going to show, with an example below, how a mass gap develops in the theory.  The self-energy function is given by $\Pi(k)$. This is quite a general behavior of the scalar field theory that happens to hold also in the non-local case. 

The propagator we obtained shows a single pole only if the function $f(\Box)$ has no zeros as is the case for an exponential form factor. This we will show explicitly in the next sub-section\footnote{A few considerations are in order about the degrees of freedom (d.o.f.) discussions of the theory, as in Ref. \cite{Kimura:2016irk,Calcagni:2018gke}. We expect a single d.o.f. in this case for the scalar field having a single component.}.

\subsubsection{Case with Gaussian Non-local Operator}



\begin{figure}[H]
\centering
\includegraphics[height=8cm,width=10cm]{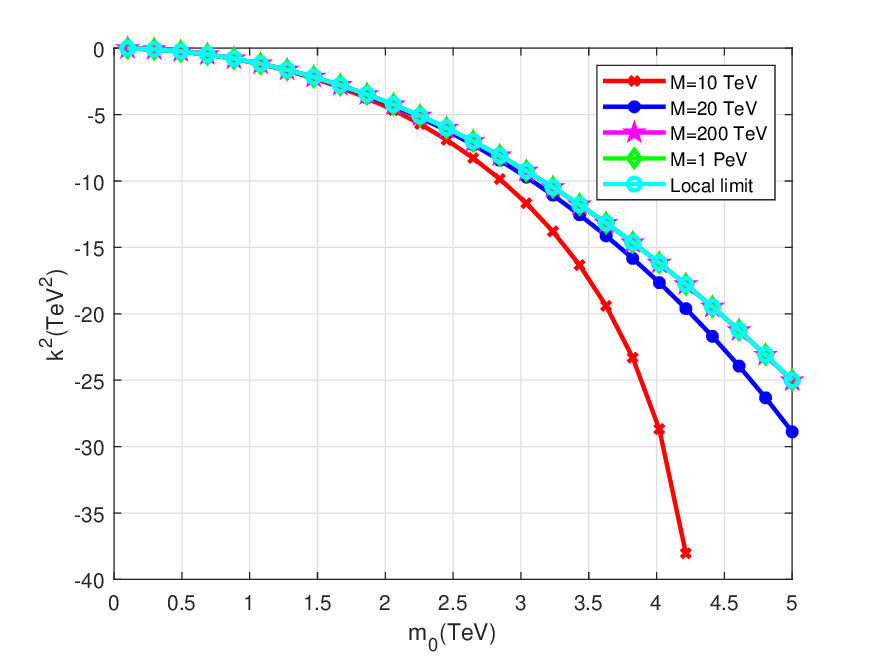}
\caption{\it Plot of the mass gap solution as a function of $m_0$. The mass gap gets damped in the UV.  
We see for $M$ being O(10 TeV), the curves do not change appreciably reaching the local limit $M\rightarrow\infty$.}\label{fig1}
\end{figure}

For the sake of completeness, we give the leading order value of the mass gap in a simple case. 
For consistency reasons with our preceding computations, we assume Euclidean metric everywhere and do not consider possible Minkowski case where it may yield tachyonic excitations that are interesting to study anyway but is beyond the scope of our current investigation in this paper. The mass gap
is obtained from the propagator in Eqn.(\ref{eq:G_lo}). If we assume $f(-k^2)=-k^2/M^2$ \footnote{We can, in principle, select a large class of entire functions instead of the Gaussian operator \cite{Edholm:2016hbt}.} 
and obtain  on-shell conditions, from the propagator equation 
\begin{equation}
\label{eq:on-shell}
(\Box+m_0^2e^{2\frac{\Box}{M^2}})\Delta(x-y)=e^{\frac{\Box}{M^2}}\delta^4(x-y)
\end{equation}
the pole is obtained at:
\begin{equation}
    k^2=\frac{M^2}{2}W(-2m_0^2/M^2)
\end{equation}
being $W(x)$ the Lambert function. It should be noticed that the Lambert function has also an infinite set of complex solutions representing nonphysical states that we do not consider here. The behaviour of the mass gap as a function of $m_0$ is presented in Fig.(\ref{fig1}). It should be remembered here that $m_0$ is a constant arising from the self-interaction part of the theory as given in Eqn.(\ref{eq:m0}).  Anyway, we emphasize that all these factors are finite due to the non-local contributions.




Finally, we just note that, in the limit $\lambda\rightarrow 0$ 
, the free propagator is properly recovered as
\begin{equation}
    \Delta_0(k)=\frac{e^{f(-k^2)}}{k^2}.
\end{equation}

For the classical theory, this completes our discussion for the non-local case, starting from known results about the local theory. Now we turn to the quantum case to see how the quantum corrections modify the scenario.


\section{Mass Gap in Quantum Case}
\label{Quantum}

The technique we work out here is based on Bender, Milton and Savage method \cite{Bender:1999ek} and we adopt it for the non-local theory. Our aim is to get a set of \textit{Dyson-Schwinger} equations in differential form. 

For the action in Eqn.(\ref{Action}), with source j, the partition function can be simply written down as
\begin{equation}
    Z[j]={\cal N}\int{\cal D}\phi e^{-S-\int d^4xje^{f(\Box)}\phi}.
\end{equation}
In order to proceed with the Green's function we average the equation of motion (\ref{eq:mot}) and divide by $Z[j]$ yielding
\begin{equation}
\label{eq:G1}
    -\Box G_1^{(j)}=-\frac{\lambda}{3!}Z^{-1}[j]\langle e^{f(\Box)}\left(e^{f(\Box)}\phi \right)^3\rangle+e^{f(\Box)}j,
\end{equation}
where we have introduced the 1-point function $G_1^{(j)}=\langle\phi\rangle/Z[j]$. We are interested in the limit $j=0$. Let us rewrite this as
\begin{equation}
    Z[j]e^{f(\Box)}G_1^{(j)}=\langle e^{f(\Box)}\phi\rangle.
\end{equation}
Derivative of this equation with respect to $j(x)$ gives
\begin{equation}
    Z[j][e^{f(\Box)}G_1^{(j)}(x)]^2+Z[j]G_2^{(j)}(x,x)=\langle (e^{f(\Box)}\phi(x))^2\rangle.
\end{equation}
Here is $G_2^{(j)}(x,y)= Z^{-1}[j]\langle e^{f(\Box)}\phi(x)e^{f(\Box)}\phi(y)\rangle$. 

Once more, performing derivation with respect to $j(x)$, one gets
\begin{eqnarray}
    &&Z[j][e^{f(\Box)}G_1^{(j)}(x)]^3+2Z[j]e^{f(\Box)}G_2^{(j)}(x,x)e^{f(\Box)}G_1^{(j)}(x)+ \nonumber \\
    &&Z[j]e^{f(\Box)}G_1^{(j)}(x)G_2^{(j)}(x,x)+Z[j]G_3^{(j)}(x,x,x)=\langle (e^{f(\Box)}\phi(x))^3\rangle
\end{eqnarray}
with $G_3^{(j)}(x,y,z)= Z^{-1}[j]\langle e^{f(\Box)}\phi(x)e^{f(\Box)}\phi(y)e^{f(\Box)}\phi(z)\rangle$. 

We can put this into Eqn.(\ref{eq:G1}) to obtain
\begin{equation}
\label{eq:G1j}
    -\Box G_1^{(j)}=-\frac{\lambda}{3!} e^{f(\Box)}\left[[e^{f(\Box)}G_1^{(j)}(x)]^3+3G_2^{(j)}(x,x)e^{f(\Box)}G_1^{(j)}(x)+G_3^{(j)}(x,x,x)\right]+e^{f(\Box)}j,
\end{equation}
By setting $j=0$, we get the Dyson-Schwinger equation for the \textit{1-point function}:
\begin{equation}
\label{eq:G1DS_0}
    -\Box G_1(x)=-\frac{\lambda}{3!} e^{f(\Box)}\left[[e^{f(\Box)}G_1(x)]^3+3G_2(x,x)e^{f(\Box)}G_1(x)+G_3(x,x,x)\right].
\end{equation}
For the scalar field, one has $G_3(x,x,x)=0$. We also notice a mass term that arises due to renormalization effects given by
\begin{equation}
\label{eq:m2r}
    \delta m_0^2=\frac{\lambda}{2!}G_2(0)
\end{equation}
where use has been made of invariance by translations. So, finally
\begin{equation}
\label{eq:G1DS}
    -\Box G_1(x)=-\delta m_0^2e^{2f(\Box)}G_1(x)-\frac{\lambda}{3!}e^{f(\Box)}[e^{f(\Box)}G_1(x)]^3.
\end{equation}
By comparing Eqns.(\ref{result_classicl}) and (\ref{eq:G1DS}), we note that the mass term is the only difference between the classical case and the quantum cases. We note that the term $\delta m_0^2$, arising purely from the quantum effects, will be present in all the Dyson-Schwinger hierarchy of equations.

Now, we take the functional derivative of eq.(\ref{eq:G1j}) with respect to $j(y)$. This will give
\begin{eqnarray}
   -\Box G_2^{(j)}(x,y) &=&
    -\frac{\lambda}{2!}e^{f(\Box)}\{[e^{f(\Box)}G_1^{(j)}(x)]^2e^{f(\Box)}G_2^{(j)}(x,y)\}
    \nonumber \\
    &&-\frac{\lambda}{2!}e^{f(\Box)}\left[G_3^{(j)}(x,x,y)e^{f(\Box)}G_1^{(j)}(x)+G_2^{(j)}(x,x)e^{f(\Box)}G_2^{(j)}(x,y)\right]
    \nonumber \\
    &&-\frac{\lambda}{3!}e^{f(\Box)}G_4^{(j)}(x,x,x,y)+e^{f(\Box)}\delta^4(x-y). 
\end{eqnarray}
Taking $j=0$ and using translation invariance, one gets
\begin{equation}
    -\Box G_2(x,y)=
    -\frac{\lambda}{2}e^{f(\Box)}\{[e^{f(\Box)}G_1(x)]^2e^{f(\Box)}G_2(x,y)\}
    -\delta m_0^2e^{f(\Box)}G_2(x,y)+e^{f(\Box)}\delta^4(x-y).
\end{equation}
This recovers again the classical equation, provided we consider the re-normalized mass in eq.(\ref{eq:m2r}) and we observe that $G_3=0$ and $G_4(0,0,x-y)=0$.

If the effect of re-normalization can be considered as a small shift in the spectrum, we can map the classical result on the quantum one and recover the spectrum of the theory at the leading order in the quantum case. 

In order to see the effect of the mass shift, we exploit the linearity of this equation and write it as an integral equation
\begin{equation}
    G_2(x-y)=G_2^{(c)}(x-y)-\delta m_0^2\int d^4zG_2^{(c)}(x-z)e^{f(\Box)}G_2(z-y)
\end{equation}
where $G_2^{(c)}$ is the classical 2-point function given by Eqn.(\ref{final_classical}), at the chosen iteration in the momenta coordinates. By a Fourier transform, it is easy to get
\begin{equation}
    G_2(k)=G_2^{(c)}(k)\frac{1}{1+\delta m_0^2e^{f(-k^2)}G_2^{(c)}(k)},
\end{equation}
where the shift $\delta m_0^2$ can then be computed by the gap equation
\begin{equation}
  \delta m_0^2=\frac{\lambda}{2!}\int\frac{d^4k}{(2\pi)^4}G_2(k).  
\end{equation}
Thus we arrived at the final summit, that is, the mass gap in the quantum field theory involving infinite-derivative non-local scalar field in the strong coupling limit.

\medskip

\section{Discussion \& Conclusion}
\label{Conclusion}

In a simple quartic scalar field theory, we showed that the mass gap and the spectrum of the states is completely changed in the non-local theory than that in the local scenario. We summarize the main findings of our paper as follows:
\begin{itemize}
    \item We showed that the \textit{Dyson-Schwinger} approach can be straightforwardly extended from the standard field theory to the infinite-derivative case, therefore 
    providing a pathway to access non-perturbative physics in the non-local theories.
    \item We studied the \textit{mass gap} generated in non-local infinite-derivative Higgs field theory and showed, as an example, for the case of non-local Gaussian operator, that the non-local scale M is responsible for no extra poles in the propagator even in the non-perturbative regime (see Eqn.(\ref{eq:on-shell})).
    \item In the limit $M \rightarrow \infty$, the mass gap becomes just like in the local theory.
    \item In the limit of the coupling $\lambda \rightarrow 0$  we recover the proper massless propagator (see Eqn.13) in the weak perturbation scenario.
    \item We notice that in the UV, beyond the scale of non-locality M, the mass gap generated in the Higgs
    gets exponentially suppressed and the theory transcends into conformal limit in that regime,
    despite the interactions are present. 
    
    
    \item For any given non-local scale, we showed that the mass gap gets exponentially suppressed with the renormalized mass in the quantum case which means that if the mass gap in the local theory
    increases the extra contribution of the effective mass gap in the non-local limit decreases.
    \item We computed the spectrum of the non-local theory, and provided a technique to evaluate, in principle, all correlation functions. We also presented presented the evaluation of the \textit{1- and 2-point-functions} in the classical and the quantum cases for the non-local field theory.
\end{itemize}
We speculate that our study and the results obtained can find straight-forwardly wide applications in modern cosmology, like one particular application for the \textit{scale free} theory comes in the context of cosmological inflation. In this regard the mass gap together with the mass-scale $M$ will play a significant role in breaking the scale invariance, as well as creating the observed density fluctuations in the cosmic microwave background radiation \cite{pdg2020}. Similarly, for dark energy models with scalar fields, it can have interesting implications in the sense that the mass gap in the theory acts as source of current cosmic acceleration but was damped in early universe due to the exponentially damping from the presence of non-local scale M in the UV, thereby offering an explanation for the fine-tuning problem but details is beyond the scope of the current manuscript.


We expect this nature on the non-local scale M to transcend a theory from non-conformal regime to scale-free,
by exponentially damping the mass gap in the UV, to be valid even for Yang-Mills theory as well.
We look to extend our analysis to a non-local Yang-Mills theory; here we point out that, due to the natural mapping between the scalar theory and Yang-Mills theory \cite{Frasca:2009yp,Frasca:2015yva}, 
we expect the same phenomena to survive also in the
non-local Yang-Mills case, i.e. an exponentially damped mass gap, but detailed computations will be taken up in another publication \cite{Frasca}.

\section{Acknowledgements}
\label{Asck}

Authors acknowledge Luca Buoninfante for several very useful suggestions on the manuscript, and also thank Anupam Mazumdar and Nobuchika Okada for comments.

\section*{Appendix: Dyson-Schwinger Equations \& Bender-Milton-Savage Technique}
\label{AppendixA}

The idea in the Bender-Milton-Savage technique to derive the Dyson-Schwinger equations \cite{Bender:1999ek} is to retain their PDE form never introducing the vertexes into the computation.

Then, we consider the partition function of a given theory, a scalar field theory to fix the ideas, into the form
\begin{equation}
    Z[j]=\int[D\phi]e^{iS(\phi)+i\int d^4xj(x)\phi(x)}.
\end{equation}
Then, we will have, for the 1P-function
\be
\left\langle\frac{\delta S}{\delta\phi(x)}\right\rangle_j=j(x)
\ee
where
\be
\left\langle\ldots\right\rangle=\frac{\int[D\phi]\ldots e^{iS(\phi)+i\int d^4xj(x)\phi(x)}}{\int[D\phi]e^{iS(\phi)+i\int d^4xj(x)\phi(x)}}
\ee
Then, we set $j=0$. The next step is to derive this equation again with respect to $j$ to compute the equation for the 2P-function. We assume the following definition of the nP-functions
\be
\langle\phi(x_1)\phi(x_2)\ldots\phi(x_n)\rangle=\frac{\delta^n\ln(Z[j])}{\delta j(x_1)\delta j(x_2)\ldots\delta j(x_n)}.
\ee
Then,
\be
\frac{\delta G_k(\ldots)}{\delta j(x)}=G_{k+1}(\ldots,x).
\ee

We now consider the simple case of a $\phi^4$ theory. Then,
\be
S=\int d^4x\left[\frac{1}{2}(\partial\phi)^2-\frac{\lambda}{4}\phi^4\right].
\ee
It is easy to see that we have to evaluate
\be
\label{eq:G_1}
\partial^2\langle\phi\rangle+\lambda\langle\phi^3(x)\rangle = j(x).
\ee
This gives
\be
Z[j]\partial^2G_1^{(j)}(x)+\lambda\langle\phi^3(x)\rangle = j(x).
\ee
From the definition of the 1P-function one has
\be
Z[j]G_1^{(j)}(x)=\langle\phi(x)\rangle.
\ee
We derive with respect to $j(x)$ obtaining
\be
Z[j][G_1^{(j)}(x)]^2+Z[j]G_2^{(j)}(x,x)=\langle\phi^2(x)\rangle.
\ee
Deriving again, this gives
\be
Z[j][G_1^{(j)}(x)]^3+3Z[j]G_1^{(j)}(x)G_2(x,x)+Z[j]G_3^{(j)}(x,x,x)=\langle\phi^3(x)\rangle.
\ee
We substitute this into Eqn.(\ref{eq:G_1}) and obtain
\be
\label{eq:G1_j}
\partial^2G_1^{(j)}(x)+\lambda[G_1^{(j)}(x)]^3+3\lambda G_2^{(j)}(0)G_1^{(j)}(x)+G_3^{(j)}(0,0)=Z^{-1}[j]j(x)
\ee
We observe that, due to renormalization, a mass term appeared
that, for $j=0$, yields the first Dyson-Schwinger equation into differential form
\be
\partial^2G_1(x)+\lambda[G_1(x)]^3+3\lambda G_2(0)G_1(x)+G_3(0,0)=0.
\ee

Next step is to derive Eqn.(\ref{eq:G1_j}) with respect to $j(y)$. This will give
\be
\begin{split}
&\partial^2G_2^{(j)}(x,y)+3\lambda[G_1^{(j)}(x)]^2G_2^{(j)}(x,y)+
\nonumber \\
&3\lambda G_3^{(j)}(x,x,y)G_1^{(j)}(x)
+3\lambda G_2^{(j)}(x,x)G_2^{(j)}(x,y)
+G_4^{(j)}(x,x,x,y)=\nonumber \\
&Z^{-1}[j]\delta^4(x-y)+j(x)\frac{\delta}{\delta j(y)}(Z^{-1}[j]).
    \end{split}
\ee
Setting $j=0$, one gets the equation for the 2P-function as
\be
\partial^2G_2(x,y)+3\lambda[G_1(x)]^2G_2(x,y)+
3\lambda G_3(0,y)G_1(x)
+3\lambda G_2(0)G_2(x,y)
+G_4(0,0,y)=
\delta^4(x-y).
\ee
This procedure can be iterated to any order giving, in principle, all the set of the Dyson-Schwinger hierarchy's equations in PDE form \cite{Frasca:2015yva}.

Extension of this technique to the non-local case is quite straightforward.

\bibliographystyle{unsrt}

\begin{thebibliography}{99}

\bibitem{Haber:1984rc}
H.~E.~Haber and G.~L.~Kane,
Phys. Rept. \textbf{117}, 75-263 (1985)
doi:10.1016/0370-1573(85)90051-1


\bibitem{Ostrogradsky:1850fid}
M.~Ostrogradsky,
Mem. Acad. St. Petersbourg \textbf{6}, no.4, 385-517 (1850)

\bibitem{Moffat:1990jj}
J.~W.~Moffat,
Phys. Rev. D \textbf{41}, 1177-1184 (1990)
doi:10.1103/PhysRevD.41.1177

\bibitem{Evens:1990wf}
D.~Evens, J.~W.~Moffat, G.~Kleppe and R.~P.~Woodard,
Phys. Rev. D \textbf{43}, no.2, 499-519 (1991)
doi:10.1103/PhysRevD.43.499

\bibitem{Tomboulis:1997gg} 
E.~T.~Tomboulis, 
hep-th/9702146.

\bibitem{Moffat:2011an}
J.~W.~Moffat,
[arXiv:1104.5706 [hep-th]].

\bibitem{Tomboulis:2015gfa} 
E.~T.~Tomboulis,
Phys.\ Rev.\ D {\bf 92}, no. 12, 125037 (2015)
[arXiv:1507.00981 [hep-th]].

\bibitem{Kleppe:1991rv}
G.~Kleppe and R.~P.~Woodard,
Nucl. Phys. B \textbf{388}, 81-112 (1992)
doi:10.1016/0550-3213(92)90546-N
[arXiv:hep-th/9203016 [hep-th]].

\bibitem{sft1}
E.~Witten, 
{\em Nucl.Phys.}~B268, p.~253, 1986.

\bibitem{sft2}
V.~A. Kostelecky and S.~Samuel,
{\em Nucl.Phys.}~B336, p.~263, 1990.

\bibitem{sft3}
V.~A. Kostelecky and S.~Samuel, 
{\em Phys.Lett.}, vol.~B207, p.~169, 1988.

\bibitem{padic1}
P.~G. Freund and M.~Olson, 
{\em Phys.Lett.}~B199, p.~186, 1987.

\bibitem{padic2}
P.~G. Freund and E.~Witten, 
{\em Phys.Lett.} B199, p.~191, 1987.

\bibitem{padic3}
L.~Brekke, P.~G. Freund, M.~Olson, and E.~Witten, 
{\em Nucl.Phys.}~B302, p.~365, 1988.

\bibitem{Frampton-padic}
P.~H. Frampton and Y.~Okada, 
 {\em Phys.Rev.}, vol.~D37, pp.~3077--3079, 1988.

\bibitem{marc}
T.~Biswas, M.~Grisaru, and W.~Siegel, 
  {\em Nucl.Phys.}, vol.~B708,
  pp.~317--344, 2005.

\bibitem{Tseytlin:1995uq} 
  A.~A.~Tseytlin,
  Phys.\ Lett.\ B {\bf 363}, 223 (1995)
  
\bibitem{Siegel:2003vt} 
  W.~Siegel,
  hep-th/0309093.

\bibitem{Calcagni:2013eua}
G.~Calcagni and L.~Modesto,
J. Phys. A \textbf{47}, no.35, 355402 (2014)
doi:10.1088/1751-8113/47/35/355402
[arXiv:1310.4957 [hep-th]].

\bibitem{Modesto:2011kw}
L.~Modesto,
Phys. Rev. D \textbf{86}, 044005 (2012)
doi:10.1103/PhysRevD.86.044005
[arXiv:1107.2403 [hep-th]].

\bibitem{Modesto:2012ga}
L.~Modesto,
[arXiv:1202.0008 [hep-th]].

\bibitem{Modesto:2015foa}
L.~Modesto, M.~Piva and L.~Rachwal,
Phys. Rev. D \textbf{94}, no.2, 025021 (2016)
doi:10.1103/PhysRevD.94.025021
[arXiv:1506.06227 [hep-th]].

\bibitem{Modesto:2017hzl}
L.~Modesto, L.~Rachwa\l{} and I.~L.~Shapiro,
Eur. Phys. J. C \textbf{78}, no.7, 555 (2018)
doi:10.1140/epjc/s10052-018-6035-2
[arXiv:1704.03988 [hep-th]].
  
\bibitem{Biswas:2014yia} 
  T.~Biswas and N.~Okada,
  Nucl.\ Phys.\ B {\bf 898}, 113 (2015)
  doi:10.1016/j.nuclphysb.2015.06.023
  [arXiv:1407.3331 [hep-ph]].
  
 \bibitem{Olive:2016xmw} 
  C.~Patrignani {\it et al.} [Particle Data Group],
  Chin.\ Phys.\ C {\bf 40}, no. 10, 100001 (2016).
  
\bibitem{Ghoshal:2017egr} 
  A.~Ghoshal, A.~Mazumdar, N.~Okada and D.~Villalba,
  Phys.\ Rev.\ D {\bf 97}, no. 7, 076011 (2018)
  doi:10.1103/PhysRevD.97.076011
  [arXiv:1709.09222 [hep-th]].
  
\bibitem{Ghoshal:2020lfd}
A.~Ghoshal, A.~Mazumdar, N.~Okada and D.~Villalba,
[arXiv:2010.15919 [hep-ph]].
  
\bibitem{Buoninfante:2018mre}
L.~Buoninfante, G.~Lambiase and A.~Mazumdar,
Nucl. Phys. B \textbf{944}, 114646 (2019)
doi:10.1016/j.nuclphysb.2019.114646
[arXiv:1805.03559 [hep-th]].
  
\bibitem{Buoninfante:2018gce}
L.~Buoninfante, A.~Ghoshal, G.~Lambiase and A.~Mazumdar,
Phys. Rev. D \textbf{99}, no.4, 044032 (2019)
doi:10.1103/PhysRevD.99.044032
[arXiv:1812.01441 [hep-th]].

\bibitem{Salvio:2020axm}
A.~Salvio,
[arXiv:2012.11608 [hep-th]].

\bibitem{Ghoshal:2018gpq}
A.~Ghoshal,
Int. J. Mod. Phys. A \textbf{34}, no.24, 1950130 (2019)
doi:10.1142/S0217751X19501306
[arXiv:1812.02314 [hep-ph]].
  
\bibitem{Biswas:2011ar} 
T.~Biswas, E.~Gerwick, T.~Koivisto and A.~Mazumdar,
Phys.\ Rev.\ Lett.\  {\bf 108}, 031101 (2012)
doi:10.1103/PhysRevLett.108.031101
[arXiv:1110.5249 [gr-qc]].
  
  
\bibitem{Biswas:2013cha} 
T.~Biswas, A.~Conroy, A.~S.~Koshelev and A.~Mazumdar,
Class.\ Quant.\ Grav.\  {\bf 31}, 015022 (2014)
Erratum: [Class.\ Quant.\ Grav.\  {\bf 31}, 159501 (2014)]
doi:10.1088/0264-9381/31/1/015022, 10.1088/0264-9381/31/15/159501
[arXiv:1308.2319 [hep-th]].
  
\bibitem{Frolov:2015bia} 
V.~P.~Frolov, A.~Zelnikov and T.~de Paula Netto,
JHEP {\bf 1506}, 107 (2015)
doi:10.1007/JHEP06(2015)107
[arXiv:1504.00412 [hep-th]].
  
  
\bibitem{Frolov:2015usa} 
V.~P.~Frolov and A.~Zelnikov
Phys.\ Rev.\ D {\bf 93}, no. 6, 064048 (2016)
doi:10.1103/PhysRevD.93.064048
[arXiv:1509.03336 [hep-th]].
  
\bibitem{Koshelev:2018hpt}
A.~S.~Koshelev, J.~Marto and A.~Mazumdar,
Phys. Rev. D \textbf{98}, no.6, 064023 (2018)
doi:10.1103/PhysRevD.98.064023
[arXiv:1803.00309 [gr-qc]].
  
\bibitem{Koshelev:2017bxd} 
A.~S.~Koshelev and A.~Mazumdar,
Phys.\ Rev.\ D {\bf 96}, no. 8, 084069 (2017)
doi:10.1103/PhysRevD.96.084069
[arXiv:1707.00273 [gr-qc]].
  
\bibitem{Buoninfante:2018xiw}
L.~Buoninfante, A.~S.~Koshelev, G.~Lambiase and A.~Mazumdar,
JCAP \textbf{09}, 034 (2018)
doi:10.1088/1475-7516/2018/09/034
[arXiv:1802.00399 [gr-qc]].
  
\bibitem{Cornell:2017irh} 
A.~S.~Cornell, G.~Harmsen, G.~Lambiase and A.~Mazumdar,
arXiv:1710.02162 [gr-qc].
  
\bibitem{Buoninfante:2018rlq}
L.~Buoninfante, A.~S.~Koshelev, G.~Lambiase, J.~Marto and A.~Mazumdar,
JCAP \textbf{06}, 014 (2018)
doi:10.1088/1475-7516/2018/06/014
[arXiv:1804.08195 [gr-qc]].

\bibitem{Buoninfante:2018stt}
L.~Buoninfante, G.~Harmsen, S.~Maheshwari and A.~Mazumdar,
Phys. Rev. D \textbf{98}, no.8, 084009 (2018)
doi:10.1103/PhysRevD.98.084009
[arXiv:1804.09624 [gr-qc]].

\bibitem{Abel:2019zou}
S.~Abel, L.~Buoninfante and A.~Mazumdar,
JHEP \textbf{01}, 003 (2020)
doi:10.1007/JHEP01(2020)003
[arXiv:1911.06697 [hep-th]].

\bibitem{Buoninfante:2020ctr}
L.~Buoninfante, G.~Lambiase, Y.~Miyashita, W.~Takebe and M.~Yamaguchi,
Phys. Rev. D \textbf{101}, no.8, 084019 (2020)
doi:10.1103/PhysRevD.101.084019
[arXiv:2001.07830 [hep-th]].
  
\bibitem{Biswas:2005qr} 
T.~Biswas, A.~Mazumdar and W.~Siegel,
JCAP {\bf 0603}, 009 (2006)
doi:10.1088/1475-7516/2006/03/009
[hep-th/0508194].
  
\bibitem{Biswas:2006bs} 
T.~Biswas, R.~Brandenberger, A.~Mazumdar and W.~Siegel,
JCAP {\bf 0712}, 011 (2007)
doi:10.1088/1475-7516/2007/12/011
[hep-th/0610274].
  
\bibitem{Biswas:2010zk} 
T.~Biswas, T.~Koivisto and A.~Mazumdar,
JCAP {\bf 1011}, 008 (2010)
doi:10.1088/1475-7516/2010/11/008
[arXiv:1005.0590 [hep-th]].

\bibitem{Biswas:2012bp} 
T.~Biswas, A.~S.~Koshelev, A.~Mazumdar and S.~Y.~Vernov,
JCAP {\bf 1208}, 024 (2012)
doi:10.1088/1475-7516/2012/08/024
[arXiv:1206.6374 [astro-ph.CO]].
  
\bibitem{Koshelev:2012qn} 
A.~S.~Koshelev and S.~Y.~Vernov,
Phys.\ Part.\ Nucl.\  {\bf 43}, 666 (2012)
doi:10.1134/S106377961205019X
[arXiv:1202.1289 [hep-th]].
  
\bibitem{Koshelev:2018rau} 
A.~S.~Koshelev, J.~Marto and A.~Mazumdar,
arXiv:1803.07072 [gr-qc].


\bibitem{Koshelev:2016vhi}
A.~S.~Koshelev, K.~Sravan Kumar and P.~Vargas Moniz,
Phys. Rev. D \textbf{96}, no.10, 103503 (2017)
doi:10.1103/PhysRevD.96.103503
[arXiv:1604.01440 [hep-th]].

\bibitem{SravanKumar:2018dlo}
K.~Sravan Kumar and L.~Modesto,
[arXiv:1810.02345 [hep-th]].

\bibitem{Koshelev:2020fok}
A.~S.~Koshelev and A.~Tokareva,
[arXiv:2006.06641 [hep-th]].

\bibitem{Koshelev:2020xby}
A.~S.~Koshelev, K.~S.~Kumar and A.~A.~Starobinsky,
Int. J. Mod. Phys. D \textbf{29}, 2043018 (2020)
doi:10.1142/S021827182043018X
[arXiv:2005.09550 [hep-th]].

\bibitem{Koshelev:2020foq}
A.~S.~Koshelev, K.~Sravan Kumar, A.~Mazumdar and A.~A.~Starobinsky,
JHEP \textbf{06}, 152 (2020)
doi:10.1007/JHEP06(2020)152
[arXiv:2003.00629 [hep-th]].

\bibitem{Buoninfante:2019swn}
L.~Buoninfante and A.~Mazumdar,
Phys. Rev. D \textbf{100}, no.2, 024031 (2019)
doi:10.1103/PhysRevD.100.024031
[arXiv:1903.01542 [gr-qc]].

\bibitem{Buoninfante:2018lnh}
L.~Buoninfante, G.~Lambiase and M.~Yamaguchi,
Phys. Rev. D \textbf{100}, no.2, 026019 (2019)
doi:10.1103/PhysRevD.100.026019
[arXiv:1812.10105 [hep-th]].
  
\bibitem{Barnaby:2007ve}
N.~Barnaby and N.~Kamran,
JHEP \textbf{02}, 008 (2008)
doi:10.1088/1126-6708/2008/02/008
[arXiv:0709.3968 [hep-th]].


\bibitem{Calcagni:2018lyd}
G.~Calcagni, L.~Modesto and G.~Nardelli,
JHEP \textbf{05}, 087 (2018)
[erratum: JHEP \textbf{05}, 095 (2019)]
doi:10.1007/JHEP05(2018)087
[arXiv:1803.00561 [hep-th]].

\bibitem{Briscese:2021mob}
F.~Briscese and L.~Modesto,
[arXiv:2103.00353 [hep-th]].


\bibitem{Calcagni:2018gke}
G.~Calcagni, L.~Modesto and G.~Nardelli,
Phys. Lett. B \textbf{795}, 391-397 (2019)
doi:10.1016/j.physletb.2019.06.043
[arXiv:1803.07848 [hep-th]].

\bibitem{Briscese:2018oyx}
F.~Briscese and L.~Modesto,
Phys. Rev. D \textbf{99}, no.10, 104043 (2019)
doi:10.1103/PhysRevD.99.104043
[arXiv:1803.08827 [gr-qc]].
 

\bibitem{Koshelev:2021orf}
A.~S.~Koshelev and A.~Tokareva,
[arXiv:2103.01945 [hep-th]].

\bibitem{Koshelev:2017tvv}
A.~S.~Koshelev, K.~Sravan Kumar and A.~A.~Starobinsky,
JHEP \textbf{03}, 071 (2018)
doi:10.1007/JHEP03(2018)071
[arXiv:1711.08864 [hep-th]].

\cite{Giacchini:2018wlf}
\bibitem{Giacchini:2018wlf}
B.~L.~Giacchini and T.~de Paula Netto,
JCAP \textbf{07}, 013 (2019)
doi:10.1088/1475-7516/2019/07/013
[arXiv:1809.05907 [gr-qc]].

\bibitem{Burzilla:2020utr}
N.~Burzill\`a, B.~L.~Giacchini, T.~d.~Netto and L.~Modesto,
[arXiv:2012.11829 [gr-qc]].

\bibitem{Salvio:2014soa}
A.~Salvio and A.~Strumia,
JHEP \textbf{06}, 080 (2014)
doi:10.1007/JHEP06(2014)080
[arXiv:1403.4226 [hep-ph]].

\bibitem{Salvio:2017qkx}
A.~Salvio and A.~Strumia,
Eur. Phys. J. C \textbf{78}, no.2, 124 (2018)
doi:10.1140/epjc/s10052-018-5588-4
[arXiv:1705.03896 [hep-th]].

\bibitem{Salvio:2018crh}
A.~Salvio,
Front. in Phys. \textbf{6}, 77 (2018)
doi:10.3389/fphy.2018.00077
[arXiv:1804.09944 [hep-th]].

\bibitem{Anselmi:2018tmf}
D.~Anselmi and M.~Piva,
JHEP \textbf{11}, 021 (2018)
doi:10.1007/JHEP11(2018)021
[arXiv:1806.03605 [hep-th]].

\bibitem{Anselmi:2018ibi}
D.~Anselmi and M.~Piva,
JHEP \textbf{05}, 027 (2018)
doi:10.1007/JHEP05(2018)027
[arXiv:1803.07777 [hep-th]].

\bibitem{Frasca:2019ysi}
M.~Frasca,
Eur. Phys. J. C \textbf{80}, no.8, 707 (2020)
doi:10.1140/epjc/s10052-020-8261-7
[arXiv:1901.08124 [hep-ph]].
	
\bibitem{Chaichian:2018cyv} 
M.~Chaichian and M.~Frasca,
Phys.\ Lett.\ B {\bf 781}, 33 (2018)
doi:10.1016/j.physletb.2018.03.067
[arXiv:1801.09873 [hep-th]].
	
\bibitem{Frasca:2017slg} 
M.~Frasca,
Nucl.\ Part.\ Phys.\ Proc.\  {\bf 294-296}, 124 (2018)
doi:10.1016/j.nuclphysbps.2018.02.005
[arXiv:1708.06184 [hep-ph]].
	
\bibitem{Frasca:2016sky} 
M.~Frasca,
Eur.\ Phys.\ J.\ C {\bf 77}, no. 4, 255 (2017)
doi:10.1140/epjc/s10052-017-4824-7
[arXiv:1611.08182 [hep-th]].

\bibitem{Frasca:2015yva} 
M.~Frasca,
Eur.\ Phys.\ J.\ Plus {\bf 132}, no. 1, 38 (2017)
Erratum: [Eur.\ Phys.\ J.\ Plus {\bf 132}, no. 5, 242 (2017)]
doi:10.1140/epjp/i2017-11563-0, 10.1140/epjp/i2017-11321-4
[arXiv:1509.05292 [math-ph]].
  
\bibitem{Frasca:2015wva}
M.~Frasca,
Eur. Phys. J. Plus \textbf{131}, no.6, 199 (2016)
doi:10.1140/epjp/i2016-16199-x
[arXiv:1504.02299 [hep-ph]].

\bibitem{Frasca:2013tma}
M.~Frasca,
Eur. Phys. J. C \textbf{74}, 2929 (2014)
doi:10.1140/epjc/s10052-014-2929-9
[arXiv:1306.6530 [hep-ph]].

\bibitem{Frasca:2012ne}
M.~Frasca,
J. Nonlin. Math. Phys. \textbf{20}, no.4, 464-468 (2013)
doi:10.1080/14029251.2013.868256
[arXiv:1212.1822 [hep-th]].

\bibitem{Frasca:2009bc}
M.~Frasca,
J. Nonlin. Math. Phys. \textbf{18}, no.2, 291-297 (2011)
doi:10.1142/S1402925111001441
[arXiv:0907.4053 [math-ph]].

\bibitem{Frasca:2010ce}
M.~Frasca,
PoS \textbf{FACESQCD}, 039 (2010)
doi:10.22323/1.117.0039
[arXiv:1011.3643 [hep-th]].

\bibitem{Frasca:2008tg}
M.~Frasca,
Nucl. Phys. B Proc. Suppl. \textbf{186}, 260-263 (2009)
doi:10.1016/j.nuclphysbps.2008.12.058
[arXiv:0807.4299 [hep-ph]].

\bibitem{Frasca:2009yp}
M.~Frasca,
Mod. Phys. Lett. A \textbf{24}, 2425-2432 (2009)
doi:10.1142/S021773230903165X
[arXiv:0903.2357 [math-ph]].

\bibitem{Frasca:2008zp}
M.~Frasca,
Int. J. Mod. Phys. E \textbf{18}, 693-703 (2009)
doi:10.1142/S0218301309012781
[arXiv:0803.0319 [hep-th]].

\bibitem{Frasca:2007uz}
M.~Frasca,
Phys. Lett. B \textbf{670}, 73-77 (2008)
doi:10.1016/j.physletb.2008.10.022
[arXiv:0709.2042 [hep-th]].

\bibitem{Frasca:2006yx}
M.~Frasca,
Int. J. Mod. Phys. A \textbf{22}, 2433-2439 (2007)
doi:10.1142/S0217751X07036427
[arXiv:hep-th/0611276 [hep-th]].

\bibitem{Frasca:2005sx}
M.~Frasca,
Phys. Rev. D \textbf{73}, 027701 (2006)
[erratum: Phys. Rev. D \textbf{73}, 049902 (2006)]
doi:10.1103/PhysRevD.73.049902
[arXiv:hep-th/0511068 [hep-th]].

\bibitem{Frasca:2005mv}
M.~Frasca,
Int. J. Mod. Phys. A \textbf{22}, 1441-1450 (2007)
doi:10.1142/S0217751X07036282
[arXiv:hep-th/0509125 [hep-th]].

\bibitem{Frasca:2005fs}
M.~Frasca,
Int. J. Mod. Phys. D \textbf{15}, 1373-1386 (2006)
doi:10.1142/S0218271806009091
[arXiv:hep-th/0508246 [hep-th]].

\bibitem{Bender:1999ek} 
C.~M.~Bender, K.~A.~Milton and V.~M.~Savage,
Phys.\ Rev.\ D {\bf 62}, 085001 (2000)
[hep-th/9907045].

\bibitem{Edholm:2016hbt}
J.~Edholm, A.~S.~Koshelev and A.~Mazumdar,
Phys. Rev. D \textbf{94}, no.10, 104033 (2016)
doi:10.1103/PhysRevD.94.104033
[arXiv:1604.01989 [gr-qc]].

\bibitem{Avramidi:2000bm}
I.~G.~Avramidi,
``Heat kernel and quantum gravity,''
Lect. Notes Phys. Monogr. \textbf{64}, 1-149 (2000)
doi:10.1007/3-540-46523-5
  
\bibitem{Pius:2016jsl}
R.~Pius and A.~Sen,
JHEP \textbf{10}, 024 (2016)
[erratum: JHEP \textbf{09}, 122 (2018)]
oi:10.1007/JHEP10(2016)024
[arXiv:1604.01783 [hep-th]].

\bibitem{Briscese:2018oyx}
F.~Briscese and L.~Modesto,
Phys. Rev. D \textbf{99}, no.10, 104043 (2019)
doi:10.1103/PhysRevD.99.104043
[arXiv:1803.08827 [gr-qc]].

\bibitem{grad} I. S. Gradshteyn and I. M. Ryzhik, {\sl Table of
Integrals, Series, and Products}, Eighth Edition, (Academic Press, 2015). Formula (0.314).

\bibitem{Kimura:2016irk}
T.~Kimura, A.~Mazumdar, T.~Noumi and M.~Yamaguchi,
JHEP \textbf{10}, 022 (2016)
doi:10.1007/JHEP10(2016)022
[arXiv:1608.01652 [hep-th]].

\bibitem{Bassett:2005xm}
B.~A.~Bassett, S.~Tsujikawa and D.~Wands,
Rev. Mod. Phys. \textbf{78}, 537-589 (2006)
doi:10.1103/RevModPhys.78.537
[arXiv:astro-ph/0507632 [astro-ph]].

\bibitem{pdg2020}
P.A. Zyla et al. (Particle Data Group), Prog. Theor. Exp. Phys. 2020, 083C01 (2020).

\bibitem{Frasca}
M.~Frasca, A.~Ghoshal
draft in preparation (2020)
  
\end{thebibliography}

\end{document}